\documentclass[final,5p,times,twocolumn]{elsarticle}

\usepackage{amsmath}
\usepackage{amssymb}
\usepackage[normalem]{ulem}

\usepackage[usenames]{color}   
\usepackage[dvipsnames]{xcolor}
\definecolor{darkgreen}{rgb}{0,0.5,0}

\biboptions{numbers,sort&compress}
 
\usepackage{xcolor}

\begin{document}
	
	\begin{frontmatter}

\title{Quantum superradiance on static black hole space-times}

\author{Visakan Balakumar}

\ead{VBalakumar1@sheffield.ac.uk}

\author{Elizabeth Winstanley}

\ead{E.Winstanley@sheffield.ac.uk}

\address{Consortium for Fundamental Physics,
	School of Mathematics and Statistics,
	The University of Sheffield,\\
	Hicks Building,
	Hounsfield Road,
	Sheffield. S3 7RH United Kingdom}

\author{Rafael P.~Bernar}

\ead{rafael.bernar@pq.cnpq.br}

\author{Lu\'is C.~B.~Crispino}

\ead{crispino@ufpa.br}

\address{Faculdade de F\'isica, 
Universidade Federal do Par\'a, 
66075-110, Bel\'em, Par\'a, Brazil}

\begin{abstract}
We study the quantum analogue of the classical process of superradiance for a massless charged scalar field on a static charged black hole space-time. 
We show that an ``in'' vacuum state, which is devoid of particles at past null infinity, contains an outgoing flux of particles at future null infinity.
This radiation is emitted in the superradiant modes only, and is nonthermal in nature. 
\end{abstract}

\end{frontmatter}

\section{Introduction}
\label{sec:intro}

In the classical phenomenon of superradiance, a wave is amplified during a scattering process, resulting in a reflected wave with greater amplitude than the incident wave \cite{Brito:2015oca}.  
One manifestation of superradiance is the scattering of low-frequency bosonic waves on a rotating black hole space-time \cite{Misner:1972kx,Press:1972zz,Chandrasekhar:1985kt}.
There is a corresponding process on a static, charged Reissner-Nordstr\"om (RN) black hole space-time \cite{Bekenstein:1973mi,Nakamura:1976nc,DiMenza:2014vpa,Benone:2015bst}, known as ``charge superradiance''. 
A charged scalar field wave is amplified upon scattering on the RN black hole if its frequency is sufficiently low. 

On rotating Kerr black hole space-times, there is a quantum analogue of the classical superradiance process \cite{Starobinsky:1973aij,Unruh:1974bw}, known as Starobinskii-Unruh radiation.
The black hole spontaneously emits particles in those modes which display classical superradiance. 
This radiation is in addition to the usual Hawking radiation \cite{Hawking:1974sw}, and is independent of the temperature of the black hole.

In this paper we study the quantum analogue of classical charge superradiance, first studied by Gibbons \cite{Gibbons:1975kk}.
As with Starobinskii-Unruh radiation, a charged black hole spontaneously emits particles in the classically-superradiant modes, resulting in nonthermal emission
\cite{Khriplovich:1999gm,Khriplovich:1999qa,Khriplovich:2002qn,Gabriel:2000mg}.
Much of the literature on this topic to date has focussed on the comparison between quantum charge superradiance and the well-known Schwinger pair-creation process \cite{Schwinger:1951nm} in a strong electric field (see, for example, \cite{Damour:1974qv,Khriplovich:1999qa,Khriplovich:1999gm,Gabriel:2000mg,Khriplovich:2002qn,Chen:2012zn,Kim:2013qoa,Johnson:2019kda} for a selection of references considering a charged scalar field on an RN black hole).
In particular, for a massive quantum field, the emission rate is suppressed by an exponential factor depending on the field mass \cite{Khriplovich:1999qa,Khriplovich:1999gm,Gabriel:2000mg,Khriplovich:2002qn},
and is negligible unless the electrostatic potential at the black hole horizon significantly exceeds the square of the mass/charge ratio of the quantum field \cite{Gibbons:1975kk}.

Here we take an alternative perspective, and consider instead a massless scalar field, so that the emission is not exponentially suppressed. We focus on the construction of quantum states and the properties of quantum expectation values as the charge of the scalar field varies. We consider a massless charged scalar field minimally coupled to the RN space-time geometry and construct natural ``in'' and ``out'' vacuum states. 
Quantum charge superradiance means that these two states are not the same, with the ``in'' vacuum containing an outgoing flux of charged particles far from the black hole.

The outline of this letter is as follows.
In Sec.~\ref{sec:classical} we review the classical process of superradiance for a charged scalar field on an RN black hole, before studying the quantum analogue of this process in Sec.~\ref{sec:quantum}.
We define our ``in'' and ``out'' vacuum states, and compute the fluxes of charge and energy emanating from the black hole. 
Our conclusions are presented in Sec.~\ref{sec:conc}.
Throughout this letter, the metric has mostly plus signature.
We use units in which $G=c=\hbar =1$ and Gaussian units for electrodynamic quantities.

\section{Classical superradiance on static black hole space-times}
\label{sec:classical}

We consider a massless charged scalar field $\Phi$ evolving on the space-time of an RN black hole, which is described by the following line element
\begin{equation}
ds^{2} = - f(r) \, dt^{2} + \left[ f(r) \right] ^{-1} dr^{2}+ r^{2} d\theta ^{2} + r^{2}\sin ^{2} \theta \, d\varphi ^{2} ,
\label{eq:RNmetric}
\end{equation}
where the metric function $f(r)$ is given by 
\begin{equation}
f(r) = 1 - \frac{2M}{r} + \frac{Q^{2}}{r^{2}} ,
\label{eq:fr}
\end{equation}
with $M$ being the mass and $Q$ the electric charge of the black hole.
If $M^{2}>Q^{2}$ (which is the only possibility we consider here), 
the metric function $f(r)$ given by (\ref{eq:fr}), has two zeros, at $r=r_{\pm }$, where
\begin{equation}
r_{\pm } = M \pm {\sqrt {M^{2}-Q^{2}}}.
\label{eq:rpm}
\end{equation}
In this case $r_{+}$ is the location of the black hole event horizon 
and $r_{-}$ is the location of the Cauchy horizon.
In this paper we restrict our attention to the region exterior to the event horizon.

The dynamics of the scalar field $\Phi$ is determined by the field equation
\begin{equation}
D_{\mu} D^{\mu } \Phi =0,
\label{eq:KG}
\end{equation}
where $D_{\mu } = \nabla _{\mu } - iqA_{\mu }$ is the covariant derivative, with $A_{\mu } $ being the electromagnetic gauge potential
$A_{\mu } = (A_{0}, 0 , 0 ,0 )$, where
\begin{equation}
A_{0} = -\frac{Q}{r} ,
\label{eq:gaugepot}
\end{equation}
and we have chosen a constant of integration so that the electromagnetic potential vanishes far from the black hole.

The scalar field modes are of the form
\begin{equation}
\phi _{\omega \ell m }(t,r,\theta ,\varphi ) 
= \frac{e^{-i\omega t}}{r} {\mathcal {N}}_{\omega }X_{\omega \ell }(r)Y_{\ell m }(\theta ,\varphi ),
\label{eq:modes}
\end{equation}
where $\ell = 0,1,2,\ldots $ is the total angular momentum quantum number, $m=-\ell, -\ell + 1, \ldots , \ell -1 , \ell $ is the azimuthal angular momentum quantum number, $\omega $ the frequency of the mode, ${\mathcal {N}}_{\omega }$ is a normalization constant   and $Y_{\ell m}(\theta ,\varphi )$ is a spherical harmonic. 
We have fixed the normalization of the spherical harmonics such that 
\begin{equation}
\int Y_{\ell m }(\theta, \varphi )Y_{\ell 'm'}(\theta, \varphi ) \, \sin \theta \, d\theta \, d\varphi = \delta _{\ell \ell'}\delta _{mm'}.
\end{equation}
We define the usual ``tortoise'' coordinate $r_{*}$ by 
\begin{equation}
\frac{dr_{*}}{dr} = \frac{1}{f(r)},
\label{eq:rstar}
\end{equation}
in terms of which the radial equation for $X_{\omega \ell }(r)$ takes the form
\begin{equation}
\left[    -\frac{d^{2}}{dr_{*}^{2}} + V_{\rm {eff}}(r) \right] X_{\omega \ell }(r) = 0, 
\label{eq:radial}
\end{equation}
where the effective potential $V_{\rm {eff}}(r)$ is
\begin{equation}
V_{\rm {eff}}(r) = 
\frac{f(r)}{r^{2}} \left[ \ell \left( \ell + 1 \right) +rf'(r) \right] - \left( \omega - \frac{qQ}{r} \right) ^{2}.
\label{eq:Veff}
\end{equation}
Near the black hole event horizon, as $r\rightarrow r_{+}$ and $r_{*}\rightarrow -\infty $, and at infinity, as $r,r_{*}\rightarrow \infty $, the effective potential $V_{\rm {eff}}$, given by (\ref{eq:Veff}), has the asymptotic values
\begin{equation}
V_{\rm {eff}}(r) \sim 
\begin{cases}
-{\widetilde {\omega }}^{2} =-\left( \omega - \frac{qQ}{r_{+}} \right) ^{2}, & r_{*} \rightarrow -\infty , \\
-\omega ^{2}, & r_{*} \rightarrow \infty ,
\end{cases}
\label{eq:Vasympt}
\end{equation}
where we have defined the quantity
\begin{equation}
{\widetilde {\omega }} = \omega - \frac{qQ}{r_{+}}.
\label{eq:tomega}
\end{equation}

A basis of solutions to the radial equation (\ref{eq:radial}) consists of the usual ``in'' and ``up'' scalar field modes, which have the asymptotic forms
\begin{subequations}
	\label{eq:inup}
	\begin{equation}
	X_{\omega \ell }^{\rm {in}}(r) = 
	\begin{cases}
	B_{\omega \ell}^{\rm {in}} e^{-i{\widetilde {\omega }}r_{*}}, & r_{*} \rightarrow -\infty ,\\
	e^{-i\omega r_{*}} + A_{\omega \ell }^{\rm {in}} e^{i\omega r_{*}}, & r_{*}\rightarrow \infty ,
	\end{cases}
	\label{eq:in}
	\end{equation}
	and
	\begin{equation}
	X_{\omega \ell }^{\rm {up}}(r) = 
	\begin{cases}
	e^{i{\widetilde {\omega }}r_{*}} +
	A_{\omega \ell}^{\rm {up}} e^{-i {\widetilde {\omega }}r_{*}}, & r_{*} \rightarrow -\infty ,\\
	B_{\omega \ell }^{\rm {up}} e^{i\omega r_{*}}, & r_{*}\rightarrow \infty ,
	\end{cases}
	\label{eq:up}
	\end{equation}
\end{subequations}
respectively. 
The ``in'' modes correspond to waves incoming from past null infinity, which are partly reflected back to future null infinity and partly transmitted down the future horizon.
The ``up'' modes correspond to waves which are outgoing near the past event horizon, partly reflected back down the future horizon and partly transmitted to future null infinity.

In addition to the ``in'' and ``up'' modes defined above, it is useful to also consider the time-reverse of these modes, denoted ``out'' and ``down'' respectively. The radial functions for these modes have the asymptotic forms
\begin{subequations}
	\label{eq:outdown}
	\begin{equation}
	X_{\omega \ell }^{\rm {out}}(r) =  X_{\omega \ell }^{{\rm {in}}* }(r) =
	\begin{cases}
	B_{\omega \ell}^{{\rm {in}}*} e^{i {\widetilde {\omega }}r_{*}}, & r_{*}, \rightarrow -\infty ,\\
	e^{i\omega r_{*}} + A_{\omega \ell }^{{\rm {in}}*} e^{-i\omega r_{*}}, & r_{*}\rightarrow \infty ,
	\end{cases}
	\label{eq:out}
	\end{equation}
	and
	\begin{equation}
	X_{\omega \ell }^{\rm {down}}(r) =  X_{\omega \ell }^{{\rm {up}}* }(r) =
	\begin{cases}
	e^{-i{\widetilde {\omega }}r_{*}} +
	A_{\omega \ell}^{{\rm {up}}*} e^{i {\widetilde {\omega }}r_{*}}, & r_{*} \rightarrow -\infty ,\\
	B_{\omega \ell }^{{\rm {up}}*} e^{-i\omega r_{*}}, & r_{*}\rightarrow \infty , 
	\end{cases}
	\label{eq:down}
	\end{equation}
\end{subequations}
respectively.
The ``out'' modes have no flux ingoing at the future event horizon, while the ``down'' modes have no outgoing flux at future null infinity. 

To find the normalization constants ${\mathcal {N}}_{\omega }$, we employ the Klein-Gordon inner product $\langle \Phi _{1}, \Phi _{2}\rangle $, defined by 
\begin{equation}
\langle \Phi _{1}, \Phi _{2}\rangle 
=   i \int _{\Sigma } \left[ 
\left( D_{\mu }\Phi _{1} \right) ^{*} \Phi _{2} - \Phi _{1}^{*}D_{\mu }\Phi _{2} 
\right] {\sqrt {-g}} \, d\Sigma ^{\mu }. 
\label{eq:innerprod}
\end{equation}
This inner product is independent of the choice of Cauchy surface $\Sigma $ over which the integral is performed. 
Using a Cauchy surface close to the union of the past event horizon and past null infinity for the ``in'' and ``up'' modes, and a Cauchy surface close to the union of the future event horizon and future null infinity for the ``out'' and ``down'' modes, we find
\begin{equation}
{\mathcal {N}}_{\omega }^{\rm {in/out}} = \frac{1}{{\sqrt {4\pi \left| \omega \right| }}},
\qquad
{\mathcal {N}}_{\omega }^{\rm {up/down}} = \frac{1}{{\sqrt {4\pi \left| {\widetilde {\omega }} \right| }}}.
\label{eq:normconsts}
\end{equation}

From the radial equation (\ref{eq:radial}) we can derive the following useful Wronskian relations:
\begin{equation}
	\omega \left[ 1- \left| A^{\rm {in}}_{\omega \ell }\right| ^{2} \right] =  ~{\widetilde {\omega }} \left| B_{\omega \ell }^{\rm {in}} \right| ^{2} ,
	\quad
	{\widetilde {\omega }} \left[ 1- \left|   A^{\rm {up}}_{\omega \ell }\right| ^{2} \right] =  ~ \omega  \left| B_{\omega \ell }^{\rm {up}} \right| ^{2} ,
	\label{eq:Wronskian1}
\end{equation}
and
\begin{equation}
	{\widetilde {\omega }} B^{\rm {in}}_{\omega \ell } =   ~ \omega B^{\rm {up}}_{\omega \ell} \,.
	\label{eq:Wronskian2}
\end{equation}
From the relations (\ref{eq:Wronskian1}) we can observe the phenomenon of charge superradiance \cite{Bekenstein:1973mi,Nakamura:1976nc,Benone:2015bst,DiMenza:2014vpa}
since a scalar field mode for which $\omega {\widetilde {\omega }}<0$ will have $| A_{\omega \ell }|^{2}>1$, and hence will be reflected with a larger amplitude than it had originally.

\begin{figure}
	\includegraphics[width=0.45\textwidth]{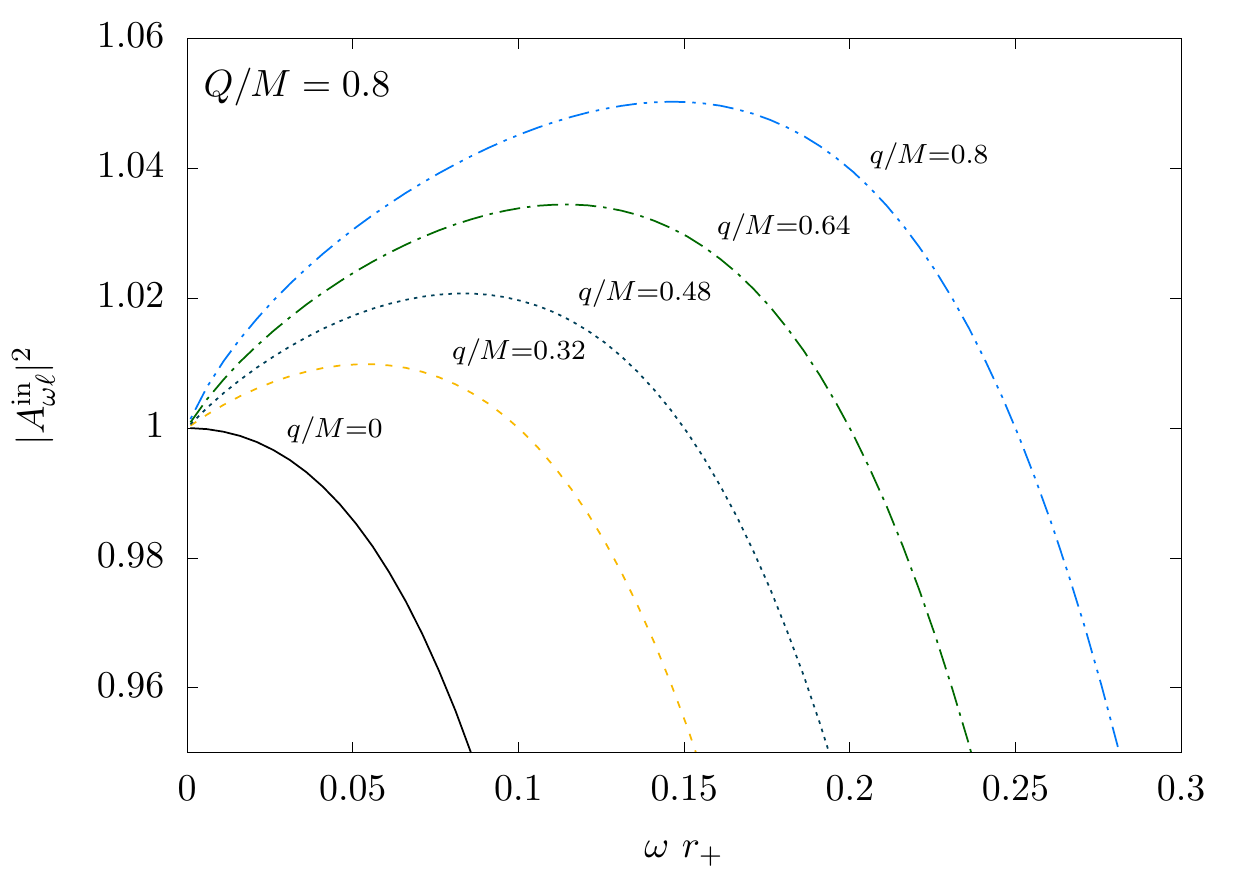}
	\caption{Reflection coefficient $|A_{\omega \ell}^{\mathrm{in}}|^2$ for the ``in'' mode with $\ell =0$ as a function of the frequency $\omega$ for some values of the scalar field charge $q$ and fixed black hole charge $Q=0.8M$.
	Superradiance occurs when $|A_{\omega \ell}^{\mathrm{in}}|^2>1$.}
	\label{fig:superradiance}
\end{figure}

This effect can also be seen in Fig.~\ref{fig:superradiance}, where we show the reflection coefficient $|A_{\omega \ell}^{\mathrm{in}}|^2$ for ``in'' modes\footnote{The Wronskian relations imply that the quantities $|A_{\omega \ell}^{\mathrm{in}}|^2$ and $|A_{\omega \ell}^{\mathrm{up}}|^2$ are equal.} with $\ell =0$ and $qQ>0$. 
Superradiance occurs for low frequency modes with $\omega <qQ/r_{+}$, when $|A_{\omega \ell}^{\mathrm{in}}|^2>1$.  
The wave amplification in this process is much larger than that observed in the superradiance of a neutral scalar field on a Kerr black hole \cite{Brito:2015oca} (see also Fig.~16 in \cite{Macedo:2012zz}).

\section{Quantum superradiance for a charged scalar field}
\label{sec:quantum}

We now turn to the quantization of the charged scalar field.
We firstly define the two quantum states of interest, and derive quantum superradiance by considering the expectation values of the current and stress-energy tensor operators.

\subsection{``In'' and ``out'' vacuum states}
\label{sec:states}

The ``in'' and ``up'' modes (\ref{eq:inup}) form an orthonormal basis of field modes.
The ``in'' modes (\ref{eq:in}) have positive norm when $\omega >0$, while the ``up'' modes (\ref{eq:up}) have positive norm when ${\widetilde {\omega }}>0$.
We therefore write the quantum field ${\hat {\Phi }}$ as the following mode sum
\begin{align}
	{\hat {\Phi }} = & ~ 
	\sum _{\ell =0}^{\infty }\sum _{m=-\ell }^{\ell } \left\{ 
	\int _{0}^{\infty } d\omega \,
	{\hat {a}}_{\omega \ell m}^{\rm {in}} \phi _{\omega \ell m}^{{\rm {in}}} +
	\int _{-\infty }^{0} d\omega \,{\hat {b}}_{\omega \ell m }^{{\rm {in}}\dagger } \phi _{\omega \ell m}^{{\rm {in}}} 
	\right. \nonumber \\ & \left. 
	+ \int _{0}^{\infty }d{\widetilde {\omega }}  \,
	{\hat {a}}_{\omega \ell m}^{\rm {up}} \phi _{\omega \ell m}^{{\rm {up}}} +
	\int _{-\infty }^{0}d{\widetilde {\omega }} \, {\hat {b}}_{\omega \ell m }^{{\rm {up}}\dagger } \phi _{\omega \ell m}^{{\rm {up}}} 
	\right\} .
	\label{eq:expBminus}
\end{align}
The expansion coefficients ${\hat {a}}$, ${\hat {b}}$ satisfy standard commutation relations (all other commutators vanish):
\begin{align}
\left[ {\hat {a}}_{\omega \ell m}^{\rm {in}}, {\hat {a}}_{\omega '\ell 'm'}^{{\rm {in}}\dagger } \right] =  & ~\delta _{\ell \ell' }\delta _{mm'} \delta (\omega - \omega ') , \qquad \omega >0,
\nonumber \\ 
\left[ {\hat {b}}_{\omega \ell m}^{\rm {in}}, {\hat {b}}_{\omega '\ell 'm'}^{{\rm {in}}\dagger } \right] =  & ~\delta _{\ell \ell' }\delta _{mm'} \delta (\omega - \omega ') , \qquad \omega <0,
\nonumber \\
\left[ {\hat {a}}_{\omega \ell m}^{\rm {up}}, {\hat {a}}_{\omega '\ell 'm'}^{{\rm {up}}\dagger } \right] =  & ~\delta _{\ell \ell' }\delta _{mm'} \delta (\omega - \omega ') , \qquad {\widetilde {\omega }}>0,
\nonumber \\ 
\left[ {\hat {b}}_{\omega \ell m}^{\rm {up}}, {\hat {b}}_{\omega '\ell 'm'}^{{\rm {up}}\dagger } \right] =  & ~\delta _{\ell \ell' }\delta _{mm'} \delta (\omega - \omega ') , \qquad {\widetilde {\omega }}<0.
\end{align} 
An ``in'' vacuum state can be defined as the state annihilated by the ${\hat {a}}^{\rm {in/up}}$ and ${\hat {b}}^{\rm {in/up}}$ operators.
We denote this state by $| {\rm {in}} \rangle $:
\begin{align}
	{\hat {a}}^{{\rm {in}}}_{\omega \ell m} | {\rm {in}} \rangle  = & ~0 ,
	\quad \omega >0, \qquad
	{\hat {b}}^{{\rm {in}}}_{\omega \ell m} | {\rm {in}} \rangle = 0,
	\quad \omega <0,
	\nonumber \\
	{\hat {a}}^{{\rm {up}}}_{\omega \ell m} | {\rm {in}} \rangle  = &~ 0 ,
	\quad {\widetilde {\omega }} >0,
	\qquad
	{\hat {b}}^{{\rm {up}}}_{\omega \ell m} | {\rm {in}} \rangle  =  0 ,
	\quad {\widetilde {\omega }}<0 .
\end{align}
The ``in'' vacuum has no particles incoming from past null infinity nor outgoing from the past horizon and hence is as empty as possible at past null infinity.

To investigate the properties of this state, it is useful to define the time-reverse of the ``in'' vacuum, namely the ``out'' vacuum $|{\rm {out}} \rangle $.
In order to construct this state, we expand the quantum scalar field ${\hat {\Phi}}$ in terms of the ``out'' and ``down'' modes (\ref{eq:outdown}):
\begin{align}
	{\hat {\Phi }} = & ~ 
	\sum _{\ell =0}^{\infty }\sum _{m=-\ell }^{\ell } \left\{ 
	\int _{0}^{\infty } d\omega \,
	{\hat {a}}_{\omega \ell m}^{\rm {out}} \phi _{\omega \ell m}^{{\rm {out}}} +
	\int _{-\infty }^{0} d\omega \,{\hat {b}}_{\omega \ell m }^{{\rm {out}}\dagger } \phi _{\omega \ell m}^{{\rm {out}}} 
	\right. \nonumber \\ & \left. 
	+ \int _{0}^{\infty }d{\widetilde {\omega }}  \,
	{\hat {a}}_{\omega \ell m}^{\rm {down}} \phi _{\omega \ell m}^{{\rm {down}}} +
	\int _{-\infty }^{0}d{\widetilde {\omega }} \, {\hat {b}}_{\omega \ell m }^{{\rm {down}}\dagger } \phi _{\omega \ell m}^{{\rm {down}}}  
	\right\} ,
	\label{eq:expBplus}
\end{align}   
where the expansion coefficients satisfy the standard commutation relations (all other commutators vanish)
\begin{align}
	\left[ {\hat {a}}_{\omega \ell m}^{\rm {out}}, {\hat {a}}_{\omega '\ell 'm'}^{{\rm {out}}\dagger } \right] =  & ~\delta _{\ell \ell' }\delta _{mm'} \delta (\omega - \omega ') , \qquad \omega >0,
	\nonumber \\ 
	\left[ {\hat {b}}_{\omega \ell m}^{\rm {out}}, {\hat {b}}_{\omega '\ell 'm'}^{{\rm {out}}\dagger } \right] =  & ~\delta _{\ell \ell' }\delta _{mm'} \delta (\omega - \omega ') , \qquad \omega <0,
	\nonumber \\
	\left[ {\hat {a}}_{\omega \ell m}^{\rm {down}}, {\hat {a}}_{\omega '\ell 'm'}^{{\rm {down}}\dagger } \right] =  & ~\delta _{\ell \ell' }\delta _{mm'} \delta (\omega - \omega ') , \qquad {\widetilde {\omega }}>0,
	\nonumber \\ 
	\left[ {\hat {b}}_{\omega \ell m}^{\rm {down}}, {\hat {b}}_{\omega '\ell 'm'}^{{\rm {down}}\dagger } \right] =  & ~\delta _{\ell \ell' }\delta _{mm'} \delta (\omega - \omega ') , \qquad {\widetilde {\omega }}<0.
\end{align}
The natural ``out'' vacuum state to define using this expansion of the quantum scalar field is then annihilated by the following ${\hat {a}}^{\rm {out/down}}$ and ${\hat {b}}^{\rm {out/down}}$ operators:
\begin{align}
	{\hat {a}}^{{\rm {out}}}_{\omega \ell m} | {\rm {out}} \rangle  = & ~0 ,
	\quad \omega >0,
	\qquad 
	{\hat {b}}^{{\rm {out}}}_{\omega \ell m} | {\rm {out}} \rangle = 0,
	\quad \omega <0,
	\nonumber \\
	{\hat {a}}^{{\rm {down}}}_{\omega \ell m} | {\rm {out}}\rangle  = & ~0 ,
	\quad {\widetilde {\omega }} >0,
	\qquad
	{\hat {b}}^{{\rm {down}}}_{\omega \ell m} | {\rm {out}} \rangle  = 0 ,
	\quad {\widetilde {\omega }}<0 .
\end{align}
The ``out'' vacuum is as empty as possible at future null infinity, and also contains no particles ingoing at the future event horizon.

\subsection{Observables}
\label{sec:exp}

We are interested in whether the ``in'' and ``out'' vacua are, in fact, identical. 
Since they have been defined in such a way that the ``out'' vacuum is the time-reverse of the ``in'' vacuum, the expectation value of the scalar field condensate $\frac{1}{2}\langle  {\hat {\Phi }}{\hat {\Phi }}^{\dagger}+ {\hat {\Phi }}^{\dagger} {\hat {\Phi }} \rangle $ will be the same in both states.
We therefore consider the expectation values of the scalar field current and stress-energy tensor, which, being tensor operators, will be able to distinguish between the two states.

The scalar field current operator ${\hat {J}}^{\mu }$ is given by 
\begin{equation}
	{\hat {{{J}}}}^{\mu } =  -\frac{i q}{16\pi } \left[ {\hat {\Phi }} ^{\dagger} \left( D^{\mu } {\hat {\Phi }} \right) + \left( D^{\mu } {\hat {\Phi }} \right){\hat {\Phi}}^{\dagger}
	- {\hat {\Phi }} \left( D^{\mu} {\hat {\Phi  }} \right) ^{\dagger } - \left( D^{\mu} {\hat {\Phi  }} \right) ^{\dagger }{\hat {\Phi }} \right] ,
	\label{eq:CalJ}
\end{equation}
and the stress-energy tensor operator ${\hat {T}}_{\mu \nu }$ takes the form
\begin{align}
	{\hat {{ {T}}}}_{\mu \nu }= & ~
	\frac{1}{4} \left\{  \left( D_{\mu }{\hat {\Phi }} \right)^{\dagger} D_{\nu }{\hat {\Phi }} 
	+ D_{\nu }{\hat {\Phi }} \left( D_{\mu }{\hat {\Phi }} \right)^{\dagger}
	+\left( D_{\nu }{\hat {\Phi }} \right)^{\dagger} D_{\mu }{\hat {\Phi }} 
	\right. \nonumber  \\ & ~ \left.
	+ D_{\mu }{\hat {\Phi }} \left( D_{\nu }{\hat {\Phi }} \right)^{\dagger}
	-	g_{\mu \nu }g^{\rho \sigma} \left[  \left(D_{\rho }{\hat {\Phi }}\right)^{\dagger}  D_{\sigma } {\hat {\Phi }}  + 
	D_{\sigma } {\hat {\Phi }} \left(D_{\rho }{\hat {\Phi }}\right)^{\dagger} 
	\right] \right\} .
	\label{eq:CalT-alt}
\end{align}
Expectation values of the current and stress-energy tensor operators in the ``in'' and ``out'' vacuum states can be written as sums over combinations of the field modes
\begin{align}
\langle {\rm {in}}| {\hat {{{J}}}}^{\mu }| {\rm {in}} \rangle =  & ~
\frac{q}{32\pi ^{2}} \sum _{\ell = 0}^{\infty } \int _{-\infty}^{\infty } d\omega 
\left(2\ell + 1 \right) \left[ j^{\mu ,{\rm {in}}}_{\omega \ell } + j^{\mu ,{\rm {up}}}_{\omega \ell}\right] , 
\nonumber \\
\langle {\rm {out}}| {\hat {{{J}}}}^{\mu }| {\rm {out}} \rangle =  & ~
\frac{q}{32\pi ^{2}} \sum _{\ell = 0}^{\infty } \int _{-\infty}^{\infty } d\omega 
\left( 2\ell + 1 \right) \left[ j^{\mu ,{\rm {out}}}_{\omega \ell } + j^{\mu ,{\rm {down}}}_{\omega \ell}\right] ,
\nonumber \\
\langle {\rm {in}}| {\hat {{{T}}}}_{\mu \nu }| {\rm {in}} \rangle =  & ~
\frac{1}{16\pi } \sum _{\ell = 0}^{\infty } \int _{-\infty}^{\infty } d\omega 
\left( 2\ell + 1 \right) \left[ t^{{\rm {in}}}_{\mu \nu ,\omega \ell } + t^{{\rm {up}}}_{\mu \nu, \omega \ell}\right] , 
\nonumber \\
\langle {\rm {out}}| {\hat {{{T}}}}_{\mu \nu }| {\rm {out}} \rangle =  & ~
\frac{1}{16\pi } \sum _{\ell = 0}^{\infty } \int _{-\infty}^{\infty } d\omega 
\left( 2\ell + 1 \right) \left[ t^{\rm {out}}_{\mu \nu ,\omega \ell } + t^{\rm {down}}_{\mu \nu ,\omega \ell}\right] , \label{eq:J_and_Tmunu}
\end{align}
where the nonzero components of the mode contributions to the expectation values are (see \cite{Balakumar} for details)
\begin{align}
j^{t,\mathrm{k}}_{\omega \ell } = & ~ -\frac{1}{r^{2} f(r)}\left|{\mathcal {N}}^{\mathrm{k}}_{\omega }\right| ^{2} \left|X^{\mathrm{k}}_{\omega \ell }(r) \right |^{2}\left( \omega - \frac{qQ}{r} \right) ,
\nonumber \\ 
j^{r,\mathrm{k}}_{\omega \ell} = & ~
 -f(r) \left| {\mathcal {N}}_{\omega }^{\mathrm{k}} \right| ^{2} \Im \left[ \frac{{X^{\mathrm{k}*}_{\omega \ell }}(r)}{r} \frac{d}{dr} \left( \frac{X^{\mathrm{k}}_{\omega \ell }(r)}{r} \right)  
\right] ,
\nonumber \\ 
t^{\mathrm{k}}_{tt,\omega \ell } = & ~
	\left| {\mathcal {N}}_{\omega }^{\mathrm{k}}\right|^{2}
\left\{ 
\left[ \frac{1}{r^{2}} \left(  \omega - \frac{qQ}{r} \right) ^{2} + \frac{\ell \left(\ell + 1 \right) f(r)}{r^{4}}  \right]  \left| X^{\mathrm{k}}_{\omega \ell }(r) \right|^{2}
\right. \nonumber \\ & ~\left. 
+ f(r)^{2}
\left| \frac{d}{dr} \left( \frac{X^{\mathrm{k}}_{\omega \ell } (r)}{r}\right) \right| ^{2} 
\right\} ,
\nonumber \\ 
	t^{\mathrm{k}}_{tr, \omega \ell  } = & ~
- 2\left( \omega - \frac{qQ}{r}\right)  \left| {\mathcal {N}}_{\omega }^{\mathrm{k}} \right| ^{2}  \Im \left[ 
\frac {{X^{\mathrm{k}*}_{\omega \ell }}(r)}{r} \frac{d}{dr}\left( \frac{X^{\mathrm{k}}_{\omega \ell}(r)}{r} \right) 
\right] ,
\nonumber \\
t^{\mathrm{k}}_{rr, \omega \ell } = & ~
\left| {\mathcal {N}}^{\mathrm{k}}_{\omega }\right|^{2}
\left\{ 
\left[ \frac{1}{f(r)^{2}r^{2}} \left(  \omega - \frac{qQ}{r} \right) ^{2} - \frac{\ell \left(\ell + 1 \right) }{r^{4}f(r)}  \right]  \left| X^{\mathrm{k}}_{\omega \ell }(r) \right|^{2}
\right. \nonumber \\ & ~ \left.
+ 
\left| \frac{d}{dr} \left( \frac{X^{\mathrm{k}}_{\omega \ell } (r)}{r}\right) \right| ^{2} 
\right\} ,
\nonumber \\
t^{\mathrm{k}}_{\theta \theta , \omega \ell } = & ~
\left| {\mathcal {N}}_{\omega }^{\mathrm{k}}\right|^{2}
\left\{ 
\frac{1}{f(r)} \left(  \omega - \frac{qQ}{r} \right) ^{2} \left| X_{\omega \ell }^{\mathrm{k}}(r) \right|^{2}
\right. \nonumber \\ & ~ \left.
- f(r)r^{2} 
\left| \frac{d}{dr} \left( \frac{X^{\mathrm{k}}_{\omega \ell } (r)}{r}\right) \right| ^{2} 
\right\} ,
\label{eq:components}
\end{align}
with $t^{\mathrm{k}}_{\varphi \varphi ,\omega \ell } =  t^{\mathrm{k}}_{\theta \theta , \omega \ell }\sin ^{2}\theta $ and $\mathrm{k}=\rm{in},\rm{up},\rm{out},\rm{down}$ labels the specific mode contribution. The symbol $\Im $ denotes the imaginary part. 

Since $X_{\omega \ell}^{\rm {out}}=X_{\omega \ell }^{{\rm {in}}*}$ and $X_{\omega \ell}^{\rm {down}}=X_{\omega \ell }^{{\rm {up}}*}$, the mode contributions $j^{t,\mathrm{k}}_{\omega \ell }$, $t^{\mathrm{k}}_{tt,\omega \ell}$, $t^{\mathrm{k}}_{rr,\omega \ell}$ and $t^{\mathrm{k}}_{\theta \theta , \omega \ell }$ are the same for the ``out'' modes as they are for the ``in'' modes, and the same for the ``down'' modes as for the ``up'' modes.
Therefore the expectation values of the corresponding components of the current and stress-energy tensor are identical in the ``in'' and ``out'' vacuum states. 
We therefore focus our attention on the remaining components, namely the fluxes $\langle {\hat {{{J}}}}^{r}\rangle$ and $ \langle {\hat {{{T}}}}_{t}^{r}\rangle $. 

\subsection{Fluxes of energy and charge}
\label{sec:fluxes}
Expectation values of the current operator in any quantum state are conserved \cite{Balakumar:2019djw}:
\begin{equation}
\nabla _{\mu }\langle {\hat {{{J}}}}^{\mu } \rangle =0.
\end{equation}
For static states as considered here, this gives
\begin{equation}
\langle {\hat {{{J}}}}^{r} \rangle = -\frac{{\mathcal{K}}}{r^{2}},
\label{eq:Jr}
\end{equation}
where ${\mathcal {K}}$ is a constant whose value depends on the quantum state under consideration.
Physically, ${\mathcal {K}}$ is the flux of charge from the black hole. 
When $\mathcal{K}$ has the same sign as the black hole charge $Q$, the black hole is losing charge.

Since there is a background electromagnetic field, expectation values of the stress-energy tensor are not conserved \cite{Balakumar:2019djw}, but instead satisfy
\begin{equation}
\nabla ^{\mu }\langle {\hat {{{T}}}}_{\mu \nu } \rangle  = 4\pi F_{\mu \nu }
\langle {\hat {{{J}}}}^{\mu } \rangle ,
\end{equation}
where $F_{\mu \nu }$ is the background electromagnetic field strength.
For static states on an RN black hole space-time, the $t$-component of this equation can be integrated to give 
\begin{equation}
\langle {\hat {{{T}}}}^{r}_{t} \rangle  = -\frac{{\mathcal {L}}}{r^{2}} +\frac{4\pi Q {\mathcal {K}}}{r^{3}},
\label{eq:Ttr} 
\end{equation}
where ${\mathcal {L}}$ is another constant depending on the particular quantum state under consideration.
Physically, ${\mathcal {L}}$ is the flux of energy from the black hole and ${\mathcal {L}}>0$ corresponds to a loss of energy by the black hole.

From the mode contributions to the expectation values (\ref{eq:components}), and using the properties $X_{\omega \ell}^{\rm {out}}=X_{\omega \ell }^{{\rm {in}}*}$ and $X_{\omega \ell}^{\rm {down}}=X_{\omega \ell }^{{\rm {up}}*}$, we have the results
\begin{equation}
\langle {\rm {out}}| {\hat {{{J}}}}^{r}|{\rm {out}} \rangle =  - \langle {\rm {in}}| {\hat {{{J}}}}^{r}|{\rm {in}}\rangle ,
\quad 
\langle {\rm {out}}|{\hat {{{T}}}}_{t}^{r}|{\rm {out}} \rangle = 
- \langle {\rm {in}}|{\hat {{{T}}}}_{t}^{r}|{\rm {in}}\rangle  ,
\end{equation}
which are to be expected since the ``out'' vacuum is the time reverse of the ``in'' vacuum.
It is therefore sufficient to study these expectation values in the ``in'' vacuum state.
It is proven in \cite{Balakumar} that these components of the current and stress-energy tensor do not require renormalization, which simplifies the computations greatly. 

We first consider the form of the expectation values $\langle {\rm {in}}| {\hat {{{J}}}}^{r}| {\rm {in}} \rangle$ and $\langle {\rm {in}}|{\hat {{{T}}}}_{t}^{r}|{\rm {in}} \rangle $ as $r\rightarrow \infty $. 
Using the form of the modes (\ref{eq:inup}), we find, as $r\rightarrow \infty $, the following leading order behaviour 
	\begin{subequations}
		\label{eq:expvaluesBminus}
\begin{align}
\langle {\rm {in}}| {\hat {{{J}}}}^{r}|{\rm {in}} \rangle & \sim  
-\frac{q}{64\pi ^{3}r^{2}}\sum _{\ell =0}^{\infty }
\int _{\min \{\frac{qQ}{r_{+}},0\}}^{\max \{ \frac{qQ}{r^{+}},0\}} d\omega 
\frac{\omega }{\left| {\widetilde {\omega }} \right| }\left(2\ell + 1 \right) \left| B^{\rm {up}}_{\omega \ell } \right| ^{2}  
,
\label{eq:JrBminus}
\\
\langle {\rm {in}}|{\hat {{{T}}}}_{t}^{r}|{\rm {in}} \rangle & \sim  -\frac{1}{16\pi ^{2}r^{2}} \sum _{\ell =0}^{\infty } 
\int _{\min \{\frac{qQ}{r_{+}},0\}}^{\max \{ \frac{qQ}{r^{+}},0\}} d\omega 
\frac{\omega ^{2}}{\left| {\widetilde {\omega }} \right| }\left(2\ell + 1 \right) \left| B^{\rm {up}}_{\omega \ell } \right| ^{2} .
\label{eq:TtrBminus}
\end{align}
\end{subequations}
These are clearly nonzero, and thus the ``in'' and ``out'' vacuum states are not the same. 
Using (\ref{eq:Jr}, \ref{eq:Ttr}), we find the constants ${\mathcal {K}}$ and ${\mathcal {L}}$ to be
\begin{subequations}
	\label{eq:KL}
\begin{align}
{\mathcal {K}} = & ~
\frac{q}{64\pi ^{3}}\sum _{\ell =0}^{\infty } 
\int _{\min \{\frac{qQ}{r_{+}},0\}}^{\max \{ \frac{qQ}{r^{+}},0\}} d\omega 
\frac{\omega }{\left| {\widetilde {\omega }} \right| } \left(2\ell + 1 \right)\left| B^{\rm {up}}_{\omega \ell } \right| ^{2}  
,
\label{eq:K} \\
{\mathcal {L}} = & ~
 \frac{1}{16\pi ^{2}} \sum _{\ell =0}^{\infty } \sum _{m=-\ell }^{\ell }
\int _{\min \{\frac{qQ}{r_{+}},0\}}^{\max \{ \frac{qQ}{r^{+}},0\}} d\omega 
\frac{\omega ^{2}}{\left| {\widetilde {\omega }} \right| }\left(2\ell + 1 \right)\left| B^{\rm {up}}_{\omega \ell } \right| ^{2}  .
\label{eq:L}
\end{align}
\end{subequations}

Both the expectation values (\ref{eq:expvaluesBminus}) involve sums over just the superradiant ``up'' modes with $\omega {\widetilde {\omega }}<0$. 
The nonzero expectation value $\langle {\rm {in}}| {\hat {{{J}}}}^{r}|{\rm {in}} \rangle$ corresponds to an outgoing flux of charge as seen by a static observer at a fixed value of the radial coordinate $r\gg r_{+}$, 
 while the nonzero expectation value $\langle {\rm {in}}|{\hat {{{T}}}}_{t}^{r}|{\rm {in}} \rangle $ represents an outgoing flux of energy as seen by that static observer.
 This is precisely the phenomenon of quantum superradiance \cite{Gibbons:1975kk}. 
 The charged black hole spontaneously emits particles in the superradiant modes.

 The fluxes (\ref{eq:KL}) contain a nonthermal distribution of particles, which is present even for extremal black holes for which the Hawking temperature vanishes.
 Since we are considering a massless charged scalar field, there is no exponential suppression of the flux, as seen in the massive case \cite{Khriplovich:1999qa,Khriplovich:1999gm,Gabriel:2000mg,Khriplovich:2002qn}.

\begin{figure}
 	\includegraphics[width=0.45\textwidth]{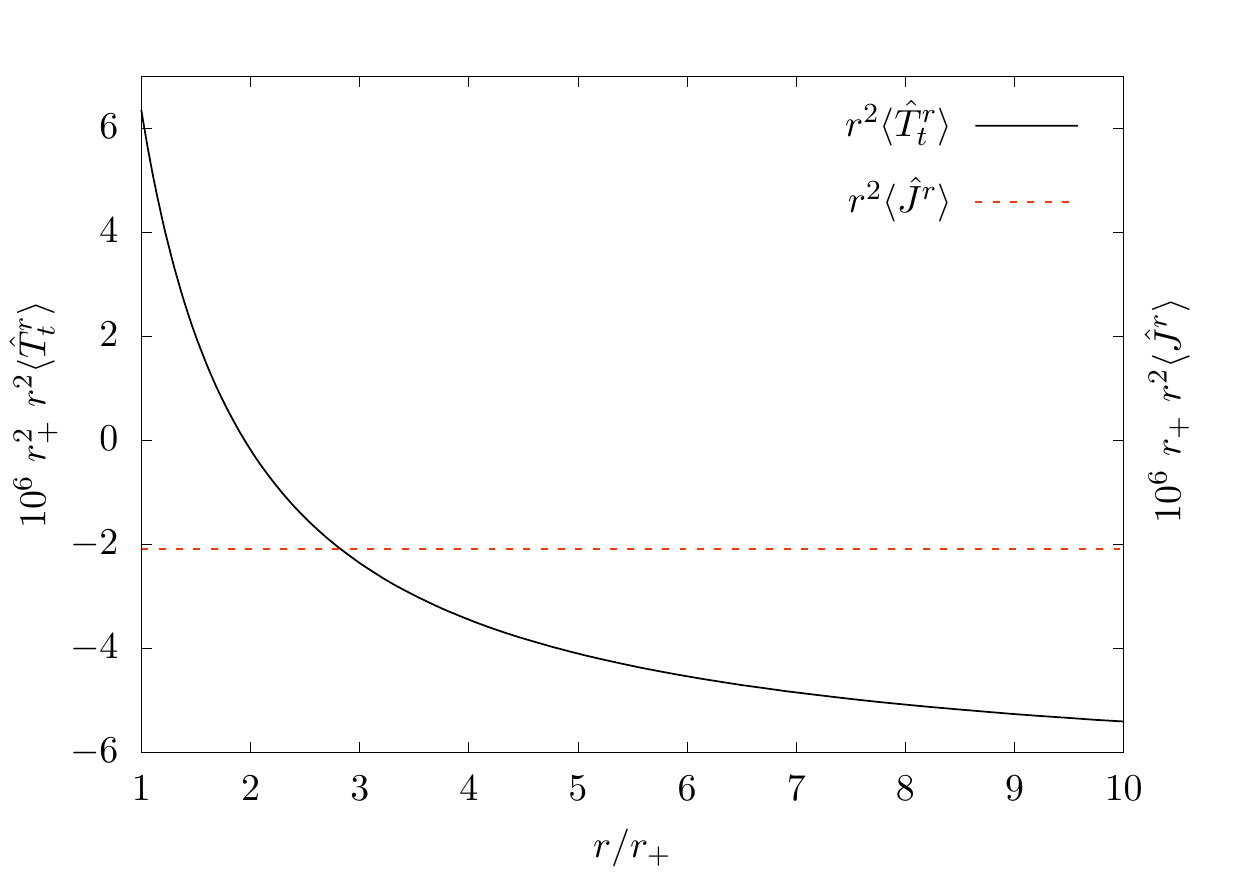}
 	\caption{Expectation values of $r^2\hat{J}^{r}$ and $r^2\hat{T}^{r}_{t}$ for the ``in'' vacuum state for $q=Q=0.8M$. The quantity $r^2\langle \hat{J}^{r} \rangle$ is constant and negative. The quantity $r^2 \langle \hat{T}^{r}_t \rangle$ is negative for sufficiently large values of $r$ but positive close to the event horizon.}
 	\label{fig:stress_rt_j_r_pB}
\end{figure}

\begin{figure}
	\includegraphics[width=0.45\textwidth]{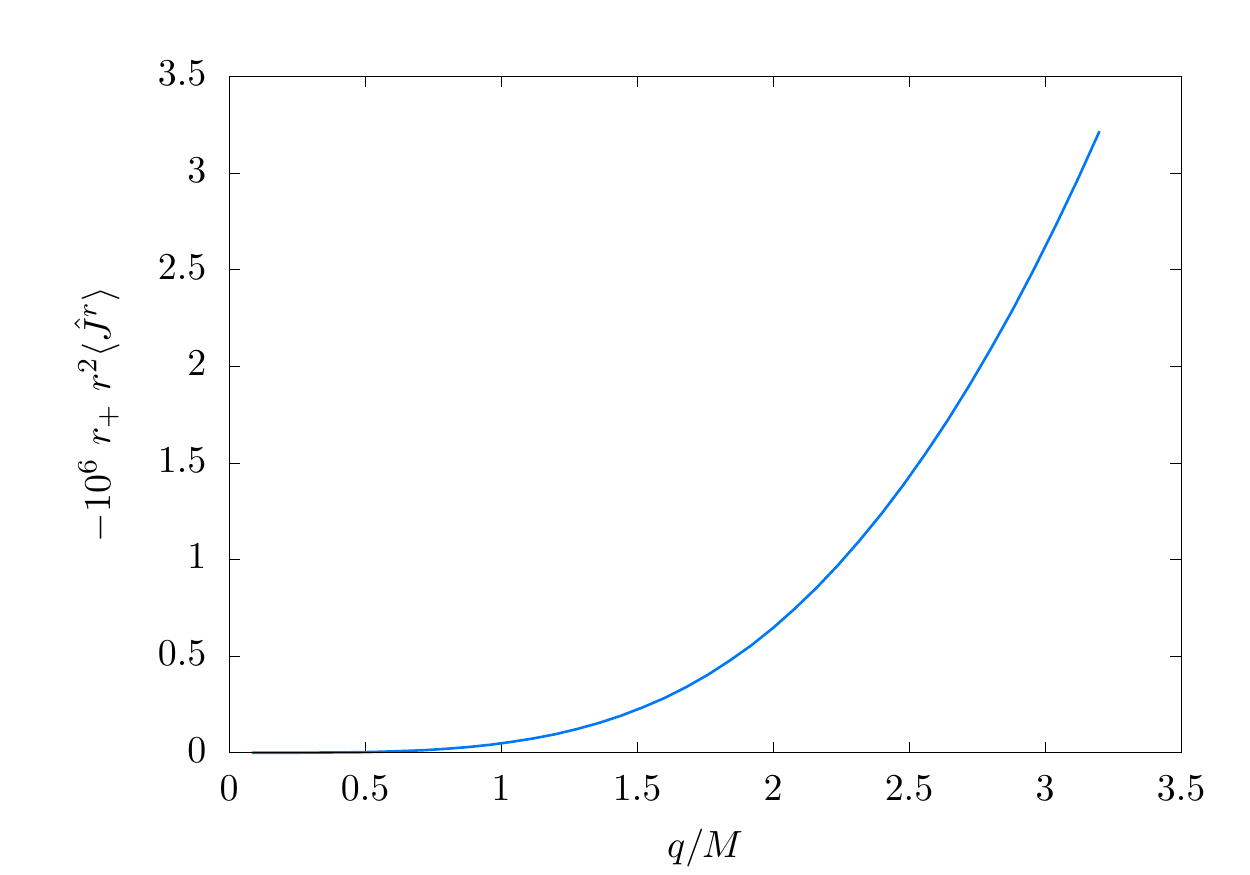}
	\caption{Expectation value of the conserved quantity $-r^{2}\langle \hat{J}^{r} \rangle$ for the ``in'' vacuum state as a function of the scalar field charge $q$, with $Q=0.8M$.}
	\label{fig:r_squared_minus_Jr_several_charges_pB}
\end{figure}

\begin{figure}
	\includegraphics[width=0.45\textwidth]{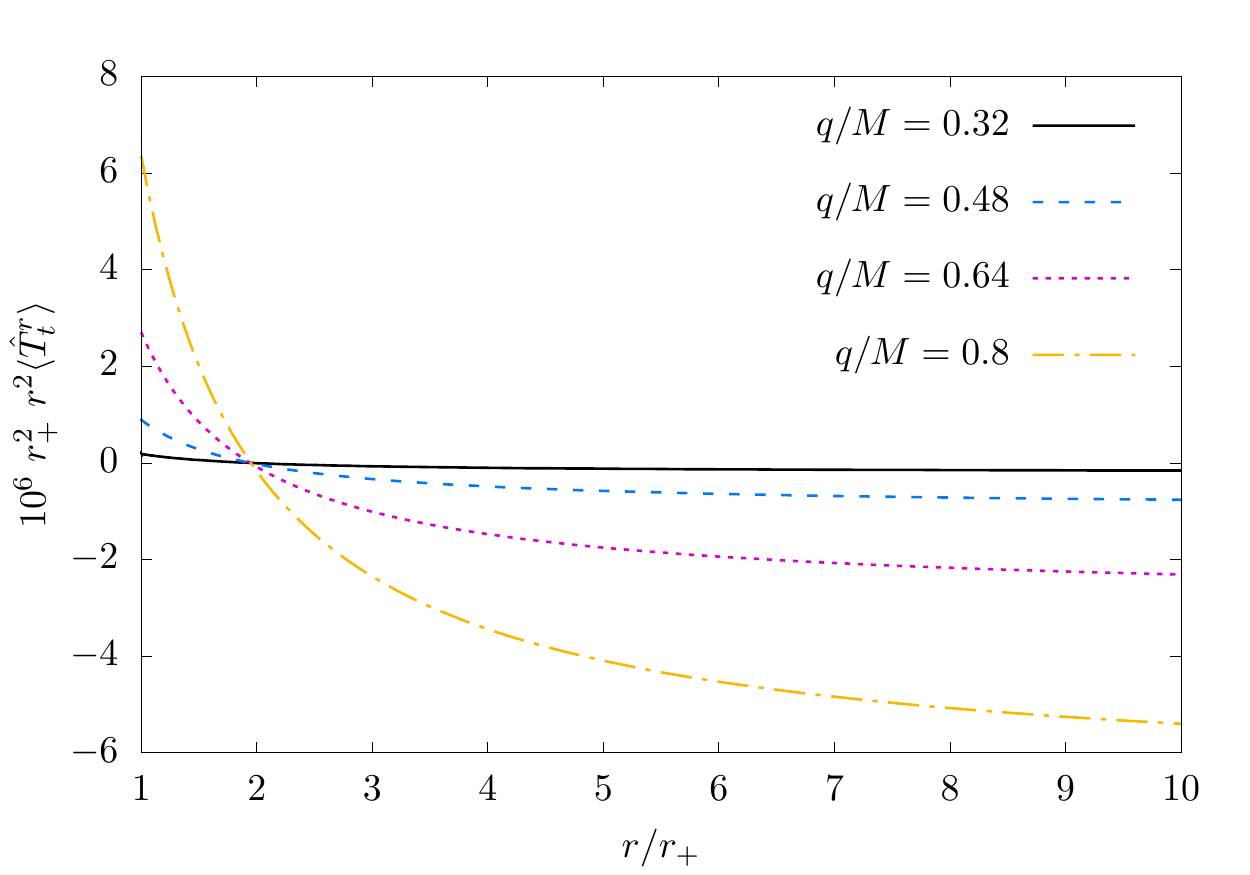}
	\caption{Expectation value $r^{2}\langle \hat{T}^{r}_{t} \rangle$ for the ``in'' vacuum state for selected values of the scalar field charge $q$, with $Q=0.8M$.}
	\label{fig:stress_rt_j_r_several_charges_pB}
\end{figure}

To calculate numerical values for the expectation values, the transmission coefficients $ \left| B^{\rm {in/up}}_{\omega \ell } \right| ^{2}$ are computed by integrating the radial equation (\ref{eq:radial}) to obtain the radial modes. These can also be inserted directly into the mode sums associated with the expectation values of $\hat{J}^{r}$ and $\hat{T}^{r}_{t}$, given in (\ref{eq:J_and_Tmunu}), as a check of our numerical results.
In Figs.~\ref{fig:stress_rt_j_r_pB}--\ref{fig:stress_rt_j_r_several_charges_pB} we display the components $r^{2}\langle {\rm {in}}| \hat{J}^{r} | {\rm {in}}\rangle $ and $r^{2}\langle {\rm {in}}|\hat{T}^{r}_{t}| {\rm {in}}\rangle $  for $Q=0.8M$ and a selection of positive values of the scalar field charge $q$.

In Fig.~\ref{fig:stress_rt_j_r_pB}, we see that, as expected, $r^{2}\langle \hat{J}^{r} \rangle $ is a constant $-{\mathcal {K}}$ (\ref{eq:Jr}, \ref{eq:K}).
Fig.~\ref{fig:r_squared_minus_Jr_several_charges_pB} shows the value of ${\mathcal{K}}$ as a function of the scalar field charge $q$ for fixed black hole charge $Q=0.8M$.   
From (\ref{eq:K}), the flux of charge ${\mathcal {K}}$ always has the same sign as the black hole charge $Q$, so that the black hole discharges due to quantum superradiance. 
As $q$ increases, it can be seen in Fig.~\ref{fig:r_squared_minus_Jr_several_charges_pB} that ${\mathcal {K}}$ increases rapidly. 

Fig.~\ref{fig:stress_rt_j_r_several_charges_pB} shows the behaviour of $r^{2}\langle {\rm {in}}|\hat{T}^{r}_{t}| {\rm {in}}\rangle $ as the scalar field charge $q$ varies.
From (\ref{eq:Ttr}), as $r\rightarrow \infty $, the quantity $r^{2}\langle {\rm {in}}|\hat{T}^{r}_{t}| {\rm {in}}\rangle $ approaches a constant $-{\mathcal {L}}$.
In Fig.~{\ref{fig:stress_rt_j_r_several_charges_pB} we see that ${\mathcal {L}}$ is always positive (as may be anticipated from (\ref{eq:L})), corresponding to a loss of energy by the black hole. 
The constant ${\mathcal {L}}$ also increases as the scalar field charge $q$ increases for fixed black hole charge.

Close to the horizon, the expectation value $\langle \hat{T}^{r}_{t}  \rangle$ is positive, due the second term in (\ref{eq:Ttr}) and the fact that $Q{\mathcal {K}}>0$.
This is in contrast to the situation for Starobinskii-Unruh radiation from a Kerr black hole \cite{Ottewill:2000qh}, for which  $\langle \hat{T}^{r}_{t}  \rangle$ has the same sign everywhere outside the event horizon. 
Therefore, at the event horizon, we find a flux of ingoing rather than outgoing energy. 
Furthermore, the magnitude of this ingoing flux at the horizon increases as the magnitude of the scalar field charge increases. 
 
The  expectation value $\langle \hat{T}^{r}_{t}  \rangle$ vanishes when $r=r_{0}$, where
\begin{equation}
r_{0}= \frac{4\pi Q{\mathcal {K}}}{{\mathcal {L}}} .
\label{eq:r0}
\end{equation}
We note that $r_0>0$ for all $Q \neq 0$. 
For all values of $q$, $Q$ studied, we find that $r_{0}>r_{+}$. 
Fig.~\ref{fig:stress_rt_j_r_several_charges_pB} seems to indicate that $r_0$ is independent of the scalar field charge $q$, for fixed black hole charge $Q$. However, there is a slight variation as one can see in Fig.~\ref{fig:r0_pB}. 
For  fixed $Q$, we find that $r_0$ increases with $q$ up to a saturation point. As $q \to 0$, the expectation value $\langle \hat{T}^{r}_{t} \rangle$ vanishes everywhere outside the black hole, which means $r_0$ is not well-defined in this limit. This is reflected in a loss of accuracy in the numerical estimation of $r_0$ for small values of $q$.

The fact that the energy flux $\langle \hat{T}^{r}_{t} \rangle$ has opposite signs close to and far from the black hole is reminiscent of the notion of an ``effective'' ergosphere \cite{Denardo:1973pyo, Denardo:1974qis, DiMenza:2014vpa}.
Inside the effective ergosphere, the energy of a charged particle can be negative as seen by an observer at infinity. 
The presence of the effective ergosphere enables a classical process of charge and energy extraction from a charged black hole, namely the charged analogue of the Penrose process.
In this process, a particle orbiting the black hole splits into two other particles with charges of opposite sign. The particle with charge of the same sign as the black hole charge escapes to infinity, whereas the other falls into the black hole, thereby effectively stealing charge from it. 
For a massless charged particle, the effective ergosphere has outermost radius given by
\begin{equation}
r_{\mathrm{ergo}}=\frac{M+\sqrt{M^2-Q^2+q^2Q^4/p_{\varphi}^2}}{1-q^2Q^2/p_{\varphi}^2},
\end{equation}
where $p_{\varphi}$ is the particle angular momentum. We note that $r_{\mathrm{ergo}}$ depends weakly on the scalar field charge $q$, in analogy with the weak dependence of $r_{0}$ on $q$ seen in Fig.~\ref{fig:r0_pB}.

\begin{figure}
	\includegraphics[width=0.45\textwidth]{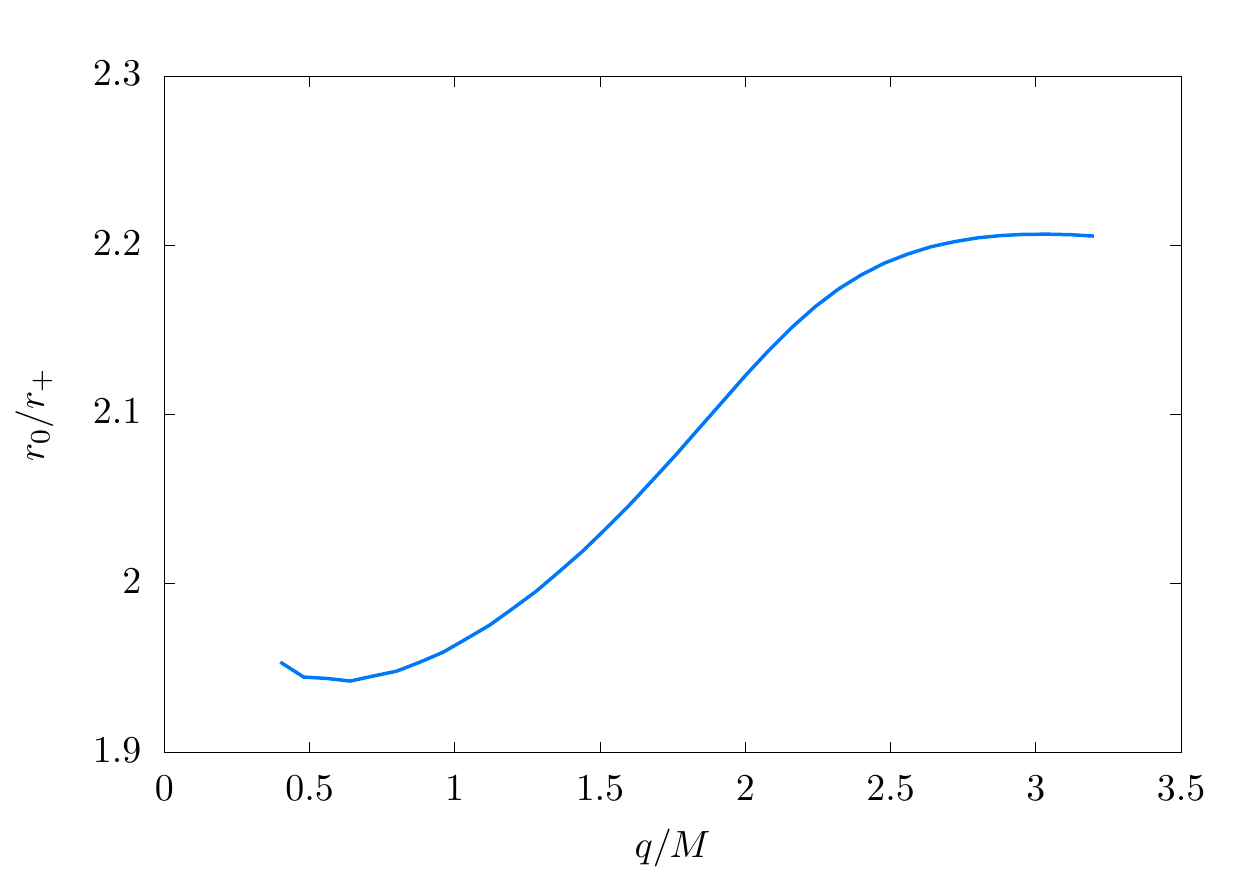}
	\caption{Radial position $r_0$, where the expectation value $\langle \hat{T}^{r}_{t} \rangle$ vanishes, for the ``in'' vacuum state, as a function of the scalar field charge $q$, with $Q=0.8M$.}
	\label{fig:r0_pB}
\end{figure}

\section{Conclusions}
\label{sec:conc}

In this letter we have studied the quantum analogue of classical superradiance for a charged scalar field on an RN black hole space-time.
For a massive field whose Compton wavelength is significantly smaller than the size of the black hole, this process is exponentially suppressed
\cite{Khriplovich:1999gm,Khriplovich:1999qa,Khriplovich:2002qn,Kim:2013qoa,Gabriel:2000mg}, so we have studied a massless scalar field, for which this exponential suppression is absent. 
We have computed the expectation values of the fluxes of charge and energy from the black hole for an ``in'' vacuum state which is as empty as possible at past null infinity.
The superradiant emission is nonthermal in nature and such that the black hole loses both charge and mass. 

As well as the ``in'' vacuum, we have also constructed an ``out'' vacuum state, which is the time-reverse of the ``in'' vacuum and is as empty as possible at future null infinity.  
The expectation values of the components of scalar field current and stress-energy tensor operators in the ``in'' and ``out'' vacua are the same, except for the fluxes of charge and energy. 
The fact that these fluxes are different means that these two vacuum states are not identical. 
Computing the fluxes is comparatively straightforward as these components of the current and stress-energy tensor operators do not require renormalization \cite{Balakumar}. 
In order to investigate the properties of the ``in'' and ``out'' vacua in more detail, we would need to examine the other components of these operators, which would require renormalization. 

Neither the ``in'' nor ``out'' vacua considered here are empty at both future and past null infinity, unlike the Boulware vacuum \cite{Boulware:1975pe} defined on a Schwarzschild black hole.
It is known that, as a consequence of quantum superradiance, there is no state empty at both future and past null infinity on a Kerr black hole \cite{Ottewill:2000qh}. Examining whether or not such a state exists for a charged scalar field on a charged black hole space-time would be an interesting extension of our work here, to which we plan to return in the near future \cite{Balakumar}.

\section*{Acknowledgments}
	V.B.~thanks STFC for the provision of a studentship supporting this work,
	and the Universidade Federal do Par\'a, for hospitality while this work was in progress.
	The work of R.P.B.~and L.C.B.C.~is financed in part by Coordena\c{c}\~ao de Aperfei\c{c}oamento de Pessoal de N\'ivel Superior (CAPES, Brazil) - Finance Code 001 and by 
	Conselho Nacional de Desenvolvimento Cient\'ifico e Tecnol\'ogico (CNPq, Brazil).
	The work of E.W.~is supported by the Lancaster-Manchester-Sheffield Consortium for Fundamental Physics under STFC grant ST/P000800/1.
	This research has also received funding from the European Union's Horizon 2020 research and innovation program under the H2020-MSCA-RISE-2017 Grant No.~FunFiCO-777740.  


\bibliography{charge-US}

\end{document}